\documentclass[useAMS, usenatbib,usegraphicx]{mn2e}
\usepackage{rotating}

\newcommand{\msun}{M$_\odot$}
\newcommand{\msunpc}{M$_\odot$ pc$^{-2}$}

\newcommand{\sigscat}{$\sigma_{\rm scat}$}

\newcommand{\sigmol}{$\Sigma_{\rm mol}$}
\newcommand{\tdep}{$\tau_{\rm dep}^{\rm CO}$}
\newcommand{\epsff}{$\epsilon_{\rm ff}^{CO}$}
\newcommand{\sigsfr}{$\Sigma_{\rm SFR}$}

\newcommand{\sigmolmeas}{${\hat \Sigma_{\rm mol}}$}
\newcommand{\sigsfrmeas}{${\hat \Sigma_{\rm SFR}}$}

\newcommand{\Xunits}{cm$^{-2}$ K$^{-1}$ km$^{-1}$ s}

\newcommand {\apgt} {\ {\raise-.5ex\hbox{$\buildrel>\over\sim$}}\ }
\newcommand {\aplt} {\ {\raise-.5ex\hbox{$\buildrel<\over\sim$}}\ }
\newcommand{\XCO}{$X_{\rm CO}$}

\newcommand{\N}{$N$}

\newcommand{\tsig}{$2\sigma$}

\title[A sub-linear and non-universal KS relationship] {Indications
  of a sub-linear and non-universal Kennicutt-Schmidt relationship}
\author[R. Shetty et al.]{Rahul Shetty$^{1}$, Brandon C. Kelly$^{2}$,
  Nurur Rahman$^{3}$\footnote{South Africa SKA Fellow}, Frank Bigiel$^{1}$, \and Alberto
  D. Bolatto$^{4}$,
  Paul C. Clark$^{1}$, Ralf S. Klessen$^{1}$, \& Lukas K. Konstandin$^{1}$  \\
$^{1}$Universit\"at Heidelberg, Zentrum f\"ur Astronomie, Institut
f\"ur Theoretische Astrophysik, Albert-Ueberle-Str. 2, 69120
Heidelberg, Germany \\
$^{2}$Department of Physics, Broida Hall, University of California,
  Santa Barbara, CA 93106, USA \\
$^{3}$Department of Physics, University of Johannesburg, Auckland Park
Campus, Johannesburg 2006, South Africa \\
$^{4}$Department of Astronomy, University of Maryland, College Park, MD, USA}

\begin{document}

\date{Accepted 2013 October 9. Received 2013 October 9; in original
  2013 June 12}

\pagerange{\pageref{firstpage}--\pageref{lastpage}} \pubyear{2013}
\maketitle

\label{firstpage}
\begin{abstract}

  We estimate the parameters of the Kennicutt-Schmidt (KS)
  relationship, linking the star formation rate (\sigsfr) to the
  molecular gas surface density (\sigmol), in the STING sample of
  nearby disk galaxies using a hierarchical Bayesian method. This
  method rigorously treats measurement uncertainties, and provides
  accurate parameter estimates for both individual galaxies and the
  entire population.  Assuming standard conversion factors to estimate
  \sigsfr\ and \sigmol\ from the observations, we find that the KS
  parameters vary between galaxies, indicating that no universal
  relationship holds for all galaxies.  The KS slope of the whole
  population is 0.76, with the \tsig\ range extending from 0.58 to
  0.94.  These results imply that the molecular gas depletion time is
  not constant, but varies from galaxy to galaxy, and increases with
  the molecular gas surface density.  Therefore, other galactic
  properties besides just \sigmol\ affect \sigsfr, such as the gas
  fraction or stellar mass.  The non-universality of the KS
  relationship indicates that a comprehensive theory of star formation
  must take into account additional physical processes that may vary
  from galaxy to galaxy.

\end{abstract}

\begin{keywords}
galaxies: ISM -- galaxies: star formation -- methods: statistical
\end{keywords}

\section{Introduction} \label{introsec}

As stars form in molecular gas, quantifying the relationship between
the star formation rate \sigsfr\ and molecular gas surface density
\sigmol\ is a prerequisite for understanding star formation in the
interstellar medium (ISM).  Most observational studies indicate a
power-law dependency
\begin{equation}
\Sigma_{\rm SFR} = a \Sigma_{\rm mol}^N,
\label{KSlaw}
\end{equation}
which is often called the ``Kennicutt-Schmidt'' (KS) relationship
\citep{Schmidt59, Kennicutt89}.  The slope \N\ is a critical parameter
of various proposed theories of star formation \citep[see e.g.][and
references therein]{MacLow&Klessen04, McKee&Ostriker07,
  Kennicutt&Evans12}.  Consequently, over the last few decades, many
observational investigations have focused on estimating the parameters
$a$ and $N$ \citep[e.g.][]{Kennicutt89, Kennicutt98, Rownd&Young99,
  Wong&Blitz02, Kennicutt+07, Bigiel+08, Bigiel+11, Leroy+08,
  Schruba+11}.

The index \N, often referred to as the KS slope, provides information
about the molecular gas depletion time \tdep, or its inverse, the star
formation efficiency over a free-fall time.  The first systematic
extra-galactic studies by \citet{Kennicutt89, Kennicutt98} estimated
\N$\sim$1.5, considering both atomic and molecular gas, suggesting
that the depletion time (efficiency) decreases (increases) with higher
total gas densities.  More recently, analyses of resolved observations
in the HERACLES \citep{Leroy+09} and STING \citep[][hereafter
R12]{Rahman+12} samples have proposed a linear KS relationship
(i.e. \N=1) for the population, often interpreted as a consequence of
constant gas depletion times (efficiency), with \tdep$\sim$2 Gyr.
These and other studies also clearly demonstrate, however, that the
large scatter in the observed trends is indicative that a single KS
relationship cannot fully describe the \sigsfr - \sigmol\ relationship
\citep[][R12]{Bigiel+08, Schruba+11, Saintonge11b, Leroy+13}.
\citet{Shetty+13}, hereafter SKB13, by performing an analysis on seven
of the HERACLES galaxies using robust Bayesian statistical methods,
reaffirm that the KS index is not constant from galaxy-to-galaxy, but
in addition find that a sub-linear relationship better describes the
data for some individual galaxies such as M51.  This suggests,
therefore, that \tdep\ increases at higher \sigmol\ for these
galaxies.

In SKB13, we demonstrated that hierarchical statistical modeling
provides significantly more accurate parameter estimates than
traditional least-squares fitting methods, which is often applied
non-hierarchically to data from all galaxies.  Moreover, Bayesian
methods allow for a rigorous treatment of measurement uncertainties
\citep{Gelman+04, Kelly07, Kruschke11}.  One question that emerges
from the SKB13 analysis is whether other observational efforts that
infer a linear or super-linear KS relationship are simply a
consequence of the chosen statistical method, or an intrinsic
relationship accurately manifested by the data.

In this work, we employ the hierarchical Bayesian method described in
SKB13 to estimate the KS parameters from the STING observational
survey \citep[][hereafter R11, and R12]{Rahman+11}.  Compared to the
\citet{Bigiel+08} HERACLES sub-sample, the STING sample contains
nearly twice as many galaxies (13 compared to 7).  In the next section
we briefly describe the STING survey.  In Section \ref{methosec} we
review the hierarchical Bayesian fitting method.  Section \ref{ressec}
presents the results of the fit on the STING data.  After a discussion
in Section 5, we conclude with a summary in Section \ref{summarysec}.

\section{Observations} \label{obssec}

We estimate the parameters of the KS relationship using the
observational data from the Survey Toward Infrared-Bright Nearby
Galaxies\footnote{http://www.astro.umd.edu/∼bolatto/STING/} (STING;
R11, R12).  Following SKB13 we denote \sigsfrmeas\ or \sigmolmeas\ as
the measured values, and \sigsfr\ or \sigmol\ as the true values of
the relevant parameters.  The STING dataset contains CARMA CO
($J=1-0$) observations of nearby disk galaxies, which indirectly
provide estimates of \sigmolmeas.  The (FWHM) resolution varies from
3\arcsec\ to 5\arcsec.  The {\it Spitzer} (MIPS) 24 \micron\ images of
these galaxies provides estimates of \sigsfrmeas, which have
$\sim$6\arcsec\ resolution.  As described in R11, we use the Nyquist
sampled maps regrided to the same 1 kpc pixel scale.  Though the MIPS
point spread function (PSF) is not a simple Gaussian, for many of the
galaxies the native resolution corresponds to substantially less than
1 kpc.  The appropriate convolution thereby minimizes the influence of
the non-Gaussian features of the PSF.  Further, we have investigated
how a more accurate convolution kernel \citep{Aniano+11} could affect
our results.  Though the KS slopes for some galaxies increase by
$\sim$0.1-0.3, the conclusions reported in the next section are not
significantly affected.  In our study, as in R12, we only consider
regions where \sigmol\ $\ge$ 20 \msunpc, where the signal to noise is
high.  For these data, the relative uncertainties of the measured CO
intensities range from $\sim 10-20 \%$.

The first two columns of Table 1 lists the galaxies and the number of
measurements in the STING sample.  We refer the reader to R11 and R12
for a more thorough description of the observations and data
reduction.

\section{Modelling Method} \label{methosec}

The hierarchical Bayesian method we employ here is described in SKB13,
where we also demonstrated its accuracy and advantage over traditional
least squares methods, such as the bisector. Here we only provide a
brief description, and refer to SKB13 for the details of the method.

Our goal is to estimate the KS parameters of each individual galaxy,
as well as the mean value of the population (group).  Fitting a single
model to combined data may conceal intrinsic variations between
individuals \citep[see also][]{Gelman&Hill07}.  Assessing the KS
parameters from observations of many regions (or beams) within a
number of galaxies is precisely a problem for which hierarchical
methods can provide robust solutions.

Furthermore, hierarchical Bayesian methods allow for a rigorous
treatment of measurement uncertainties.  For example, the \sigsfrmeas\
and \sigmolmeas\ may not be accurate due to noise and possible errors
in the assumed conversion factors.  By performing Markov Chain Monte
Carlo simulations (MCMC), the analysis explores the true value of
\sigsfr\ and \sigmol\ given the measurement \sigsfrmeas\ and
\sigmolmeas\ and their uncertainties.

After a large number of MCMC steps, the outcome of the Bayesian
analysis, or the {\it posterior}, is a probability distribution
function (PDF) for each unknown parameter, including those defining
the KS relationship.  Therefore, the posterior provides PDFs of
plausible values for the parameters, thoroughly accounting for the
uncertainties in the measurements or any other quantity, such as the
conversion factors, required in the modelling.

\section{Results} \label{ressec}

\subsection{The KS relationship of the STING galaxies}

The quantities \sigsfrmeas\ and \sigmolmeas\ are not measured
directly.  Instead, we follow the conventional practice and infer
these quantities from CO and IR observations using fixed conversion
factors.  The \XCO\ factor directly provides \sigmolmeas\ from the CO
observations, and we use the standard Galactic value, \XCO\ =
$2\times10^{20}$ \Xunits, converted to \msunpc\ units, including a
factor of 1.4 to account for the contribution from He \citep[see][and
references therein]{Bolatto+13}.  We have confirmed that allowing
\XCO\ to vary freely within a suitable range
\citep[e.g.][]{Shetty+11a, Shetty+11b} does not change our results.
To compute the star formation rate from the 24 \micron\ fluxes, we
utilize the conversion factor described by \citet{Calzetti+07}:
SFR(\msun\ yr$^{-1}$) = 1.27$\times10^{−38}[\nu$L$_{24}$(erg
s$^{−1}$)]$^{0.8850}$.

The model requires suitable estimates of the uncertainties.  As the
largest sources of error are likely the conversion factors, we choose
uncertainties of 25\% and 50\% for \sigmolmeas\ and \sigsfrmeas,
respectively.  These are possibly larger than the true uncertainties,
but we prefer to be conservative in our error estimates.

\begin{figure*}
\includegraphics*[width=170mm]{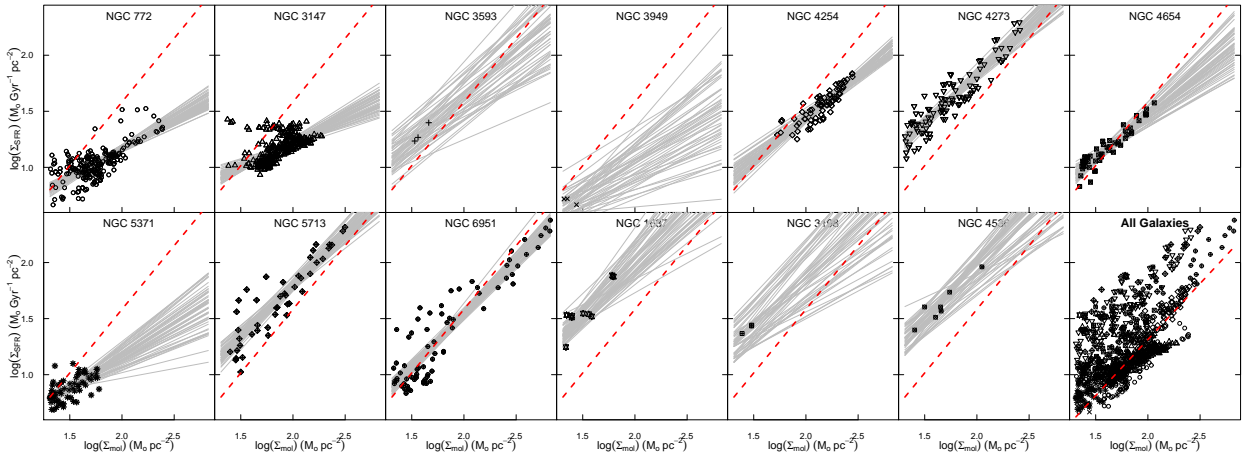}
\caption{Star formation rates and surface densities of 13 STING
  galaxies.  Black symbols show the data, and gray lines show 50
  random draws from the posterior.  For reference, (red) dashed lines
  shows the bisector fit with \N=1.1 to all data (shown together in
  last panel).}
\label{alldat}
\end{figure*}

Figure \ref{alldat} shows \sigsfrmeas\ and \sigmolmeas\ for thirteen
galaxies, as well as the ensemble in the last panel.  The gray lines
are fifty random draws from the posterior, depicting representative
results of the hierarchical Bayesian fit.  For comparison, the red
dashed line is the (non-hierarchical) bisector result $N=1.1$ on all
datapoints (shown in the last panel).  Clearly, for many galaxies the
linear bisector result cannot reproduce the \sigsfr\ - \sigmol\ trend.
For the galaxies with the largest deviations from the bisector, such
as NGC 772, NGC 3147, and NGC 4254, the data strongly evinces a {\it
  sub-linear} KS relationship.

Table 1 shows the Bayesian parameter estimates of the KS relationship
(with $A = \log a$) for all individuals and the population.  Columns 3
- 7 present the KS parameter estimates for each individual galaxy,
including the \tsig\ extent and the scatter \sigscat\ about the
regression.  The last row indicates the parameter estimates for the
population.  For five of the thirteen galaxies, the data is
inconsistent with a linear KS relationship at the \tsig\ level.
Further, the mean slope of the ensemble is estimated to be between
0.58 and 0.94, favoring a sub-linear KS relationship with 95\%
confidence.  Similar to the results from SKB13, there is significant
galaxy-to-galaxy variation, indicating that no single KS relationship
provides an accurate fit for all galaxies.

Since the posterior contains PDFs for the unknown parameters for each
individual, we can take differences in \N\ to ascertain whether the
galaxies have similar KS relationships.  For NGC 0772 and NGC 4273,
for instance, this PDF contains no values close to zero, revealing
that these galaxies have different KS relationships.

We can also estimate, generically, how many galaxies would have
sub-linear KS slopes.  Though we have find that five individual
galaxies from the STING sample have sub-linear slopes at 95\%
confidence, the total number of galaxies with sub-linear slopes
predicted by the posterior may different.  At the end of the MCMC
analysis, we can evaluate the PDF of the number of galaxies with
sub-linear slopes.  The Bayesian results indicate that at 95\%
confidence at least 9 galaxies will have sub-linear KS relationships.

\subsection{The molecular gas depletion time of STING galaxies}

The gas depletion time is defined as
\begin{equation}
\tau_{\rm dep}^{\rm CO}=\Sigma_{\rm mol}/\Sigma_{\rm SFR},
\label{tdepeqn}
\end{equation}
where we include ``CO'' in the superscript to emphasize that \tdep\
corresponds to the depletion time of the gas traced by CO.
\begin{figure*}
\includegraphics*[width=180mm]{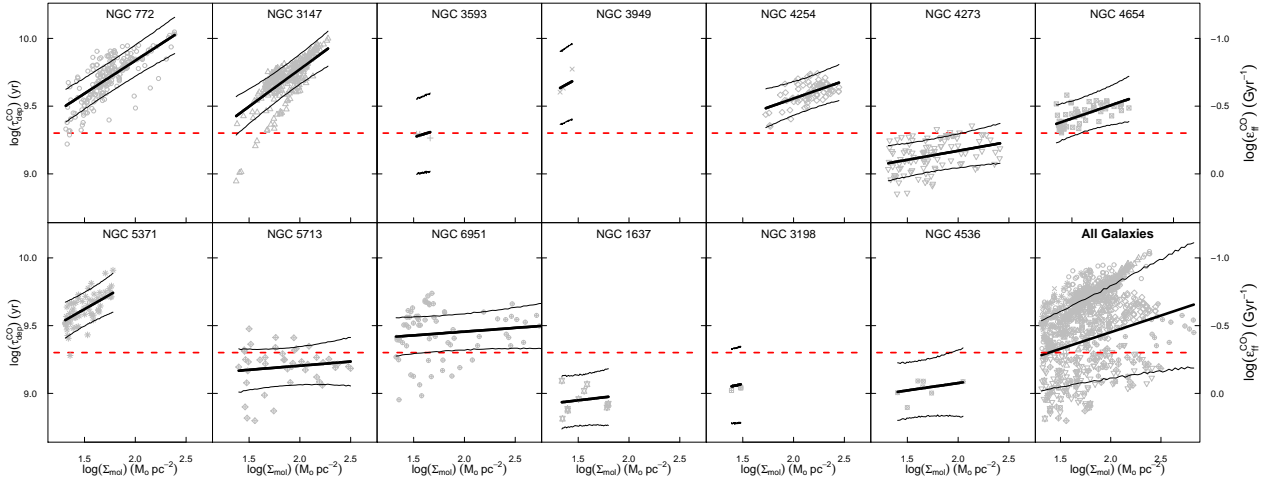}
\caption{Depletion time \tdep\ and surface density of STING Galaxies.
  Points indicate the directly measured values.  Solid line is the
  median of the Bayesian estimate, and thin lines mark the \tsig\
  interval.  The red dashed line indicates \tdep=2 Gyr.  The
  efficiency per free fall time (\epsff) is marked on the right
  ordinate.}
\label{tdepfig}
\end{figure*}
Figure \ref{tdepfig} shows \tdep\ using the measured values of
\sigsfrmeas\ and \sigmolmeas\ in Equation \ref{tdepeqn} directly, as
well as the Bayesian fit with the \tsig\ uncertainty range.  The
Bayesian model predictions for each galaxy at four values of
\sigmol=50, 100, 150, and 200 \msunpc\ are provided in Table 1.

As with the KS relationship itself in Figure \ref{alldat}, there is no
single \tdep\ that holds for all galaxies.  Further, for those
galaxies with a strongly sub-linear relationship, \tdep\ clearly {\it
  increases} with increasing gas surface density.

For instance, for NGC 772 where the median \N=0.51, the median \tdep\
varies from \aplt 5 Gyr at \sigmol=50 \msunpc, to \apgt 9 Gyr at
\sigmol=200 \msunpc.  Altogether, a constant value of \tdep=2 Gyr can
be ruled out at 95\% for all \sigmol $\ge$ 50 \msunpc\ for this
galaxy.  Note that some galaxies are consistent with a constant
depletion time.  NGC 5713, for example, has \tdep$\approx$2$\pm$1 Gyr.

\section{Caveats}

There are a number of caveats and assumptions that enter our analysis.
Missing flux intrinsic to interferometric observations, uncertainties
other than random uncorrelated errors, and variable conversion factors
are all likely to affect any parameter estimates.

As noted by R11, large scale diffuse emission may not be detected by
interferometers.  Single-dish observations are needed for recovering
this component \citep[e.g.][]{Helfer+03}.  This diffuse emission will
certainly have an impact on any estimated KS relationship, as it will
tend to increase the flux more in the low density regime, steepening
the estimated slope.  For the one galaxy for which we do have single
dish data, NGC 4254, the impact of correcting for missing flux results
in a slope increase of less than 0.1.  Fully accounting for large
scale emission may therefore impact the KS parameter estimates.  In
our analysis we have attempted to minimize the effect of diffuse
emission by only considering CO bright regions (with \sigmol\ $\ge$ 20
\msunpc).  In principle, for completeness single-dish observations
would be required.

In the hierarchical modelling, we have only considered random
statistical uncertainties.  However, we have been very conservative
and have adopted rather ``wide'' priors.  They are larger than the
noise estimates for the CO and IR fluxes.  This implicitly accounts
for unknown sources of errors.  For example, there are gain and noise
uncertainties, which we have not treated separately.  As these errors
are probably uncorrelated, using a single prior for the overall
uncertainty likely does not unduly affect the results.  However, two
possible sources of correlated uncertainties are the conversion
factors from the observed intensities to \sigsfr\ and \sigmol.  In our
analysis, if (some fraction of) the uncertainties are assigned to the
conversion factors, then they are uncorrelated.  However, as discussed
in SKB13 these factors may correlated, and in principle hierarchical
models are very well suited for investigating such correlations.  We
note that we have explored the possibility of correlations between the
\XCO\ and the conversion between 24 \micron\ and \sigsfr, but the
current STING dataset does not provide any strong statistical evidence
favoring correlated conversion factors.  Future efforts with larger
datasets and a more detailed treatment of individual contributions to
the overall uncertainties may further reveal such correlations and
their impact on the KS relationship.

Finally, we stress again that we have followed conventional practice
and employed standard conversion factors.  As such we are essentially
modelling the correlations between the CO and 24 \micron\ intensities.
Note that though we have employed a non-linear relationship between
\sigsfr\ and the 24 \micron\ intensities, the KS slope estimates are
only marginally affected (by $\sim 0.1$) when we employ a linear
conversion.  Our analysis of the STING dataset indicates that to 95\%,
the the mean KS index of the ensemble is consistent with a sub-linear
KS relationship when assuming commonly-used conversion factors, with
significant variation between galaxies.  Again, future efforts
assessing the possibility of variable conversion factors will further
shed light on the intrinsic relationship between \sigmol\ and \sigsfr.

\section{Summary \& Discussion}\label{summarysec}

We have applied a hierarchical Bayesian fitting method to the STING
sample of nearby galaxies at 1 kpc scales for estimating the KS
parameters.  Our main results are as follows:

1) The KS parameters vary from galaxy to galaxy.  The median slope
estimate ranges from as low as 0.43 (NGC 3147) to as high as 0.95 (NGC
6951).  The range in slopes of the STING sample is consistent with
that found from the SKB13 analysis of a sub-sample of 7 galaxies
from the HERACLES survey \citep{Bigiel+08}.

2) The mean value of the KS slope is sub-linear, with the median of
the PDF falling at 0.76.  Further, the data for 5 galaxies yields
N$<$1 with 95\% confidence, although considering the effects of
missing single-dish data and the 24 um PSF it is possible that 3 of
these galaxies can include \N=1 within the 95\% confidence boundary.

3) For galaxies with sub-linear relationship, assuming no dramatic
changes to the conversion factors with environment would imply an
increasing \tdep\ at higher \sigmol.  As the KS slope is not constant,
the value of \tdep\ at a given \sigmol\ also varies depending on the
galaxy.  For instance, for \sigmol=100 \msunpc, \tdep\ varies from
\aplt 1 to \apgt 9 Gyr.  Equivalently, the star formation efficiency
per free-fall time decreases with increasing CO luminosity.

These results stand in contrast with the idea of a linear relationship
between \sigsfr\ and \sigmol, or quantities commonly used as
indicators for SFR and H$_2$.  There are two primary reasons for the
discrepancy.  As we discussed in SKB13, by pooling all data together
intrinsic variations between galaxies may be veiled, with the outcome
being dominated by those galaxies with the tightest KS relationship,
and with the largest number of datapoints.  Second, the bisector is a
statistical measure that is difficult to interpret, because a slope of
unity can result from different scenarios, including those without any
correlation between the predictor and response \citep[see
also][]{Isobe+90}.  These results also suggest that no single \tdep\
(e.g. of around 2 Gyr) can accurately describe the star formation
timescale of all galaxies, confirming recent results by
\citet{Leroy+13} indicating large range in \tdep\ spanning over 0.3
dex.  Their analysis indicates that the different gas depletion times
relate to the variation in the metallicity and dust-to-gas ratio.

A sub-linear KS relationship is especially evident for NGC 772, NGC
3147, NGC 4254, and NGC 5371 (Fig. 1).  Accordingly, these galaxies
have the highest \tdep, which clearly increases with \sigmol\
(Fig. 2).  A thorough investigation of other physical properties of
these galaxies may reveal the underlying causes for a sub-linear KS
relationship, and lead to a more robust understanding about the
variations in the star formation properties between galaxies.

The significant variation in the KS parameters between galaxies
indicates that \sigsfr\ depends on other physical properties besides
just \sigmol.  For instance, the relative effects of the gas
fractions, magnetic fields, metallicity, and/or stellar mass may have
stronger influence on the \sigsfr\ than \sigmol.  In fact,
\citet{Shi+11} demonstrate a tighter correlation between \sigsfr\ with
the stellar mass, compared to \sigmol.  \citet{Leroy+13} also find
strong evidence that the KS relationship varies between galaxies as
well as between the galactic centers and outer disk regions.  Taken
these results together, \sigsfr\ needs to be assessed in the context
of other physical properties besides just \sigmol.

The result of a mean sub-linear KS relationship may simply suggest
that on average, CO is not a direct tracer of star formation activity
\citep[e.g.][]{Gao&Solomon04}.  One possible interpretation is that CO
is abundant away from star forming cores
\citep[e.g.][]{Glover&Clark12c}.  Similarly, the increasing \tdep\
with \sigmol\ may be due to the presence of excited CO in the diffuse
or non-star-forming ISM \citep[e.g.][]{Liszt+10, Pety+13}.  For
instance, towards the centers of galaxies the ISM conditions may be
conducive for CO formation, as the higher overall ambient densities
may lead to effective CO self-shielding \citep[e.g.][]{Sandstrom+13}.
Star formation, on the other hand, may require even higher densities,
so that there may not be a one-to-one correlation between CO emission
and star formation.

This interpretation, however, depends on all the other assumptions
that enter the modeling.  An important one that requires further
investigation is that CO and IR linearly trace \sigmol\ and \sigsfr.
In fact, other efforts deducing a super-linear KS relationship assume
that some fraction of the 24 \micron\ emission may be due to a
population of older stars \citep{Kennicutt+07, Liu+11}.  Moreover,
towards the dense nuclear regions of galaxies, IR emission may be
optically thick, requiring further modification to the IR-to-SFR
conversion \citep[][]{Hayward+11}.  With larger datasets, it will be
possible to assess the conversion factors in various environments,
which may lead to additional constraints on the KS relationship.

Accurate theoretical models of star formation should be able to
describe the variations in the KS parameters we find here, likely due
to numerous environmental conditions.  Analysis of multi-wavelength
observations will be needed to fully assess such models.  Hierarchical
models are suitable for handling the multi-wavelength datasets to
sample the multi-dimensional parameter space germane to such complex
models.  With additional datasets, and with expanded hierarchical
models, we can investigate the relationship between the star formation
rate and other physical properties of the ISM, possibly leading to
further insights into the physical drivers of the KS relationship.

\begin{sidewaystable*}
\setcounter{table}{1}
\vspace{-15cm}
{\bf Table 1.} Bayesian estimated parameters for the STING galaxies \\
 \centering
  \begin{tabular}{ccccccccccc}
  \hline
  \hline
Subject & \# Datapoints & $A$ & $2\sigma_A$ & $N$ & $2\sigma_N$ & \sigscat & \tdep(\sigmol=50)$^1$ & \tdep(\sigmol=100)$^1$ & \tdep(\sigmol=150)$^1$ &  \tdep(\sigmol=200)$^1$  \\
\hline
1.  NGC  772 & 217 & 0.14070 & [$-$0.09, 0.37] & 0.51 & [0.38, 0.64] & 0.05 & 3.8, 4.9, 6.3 & 5.3, 6.9, 8.9 & 6.3, 8.4, 11.1 & 7.1, 9.6, 13.0 \\
2.  NGC 3147 & 298 & 0.36924 & [0.17, 0.64] & 0.43 & [0.28, 0.53] & 0.05 & 3.1, 4.0, 5.1 & 4.6, 6.0, 7.5 & 5.8, 7.5, 9.7 & 6.8, 8.9, 11.6 \\
3.  NGC 3593 &   3 & 0.06693 & [$-$0.48, 0.61] & 0.77 & [0.43, 1.12] & 0.06 & 1.0, 2.1, 4.3 & 1.1, 2.4, 5.5 & 1.1, 2.7, 6.6 & 1.1, 2.8, 7.5 \\
4.  NGC 3949 &   3 & 0.09412 & [$-$0.61, 0.40] & 0.59 & [0.23, 0.96] & 0.06 & 2.8, 6.2, 13.4 & 3.1, 8.1, 21.0 & 3.2, 9.6, 27.4 & 3.5, 10.8, 34.6 \\
5.  NGC 4254 &  79 & 0.02614 & [$-$0.42, 0.44] & 0.74 & [0.52, 0.92] & 0.06 & 2.1, 3.0, 4.2 & 2.7, 3.6, 4.8 & 3.0, 4.0, 5.3 & 3.2, 4.3, 5.8 \\
6.  NGC 4273 & 103 & 0.09421 & [$-$0.16, 0.28] & 0.87 & [0.76, 1.02] & 0.06 & 1.1, 1.5, 1.8 & 1.1, 1.6, 2.0 & 1.1, 1.6, 2.1 & 1.2, 1.6, 2.2 \\
7.  NGC 4654 &  53 & 0.07223 & [$-$0.44, 0.26] & 0.77 & [0.57, 1.00] & 0.06 & 2.1, 2.9, 3.9 & 2.4, 3.4, 4.8 & 2.4, 3.7, 5.5 & 2.5, 3.9, 6.2 \\
8.  NGC 5371 &  65 & 0.03740 & [$-$0.36, 0.42] & 0.56 & [0.31, 0.82] & 0.06 & 3.8, 5.1, 7.0 & 4.6, 6.9, 10.4 & 5.0, 8.2, 13.6 & 5.4, 9.4, 16.4 \\
9.  NGC 5713 &  44 & 0.09586 & [$-$0.41, 0.22] & 0.95 & [0.78, 1.11] & 0.06 & 1.1, 1.5, 2.2 & 1.1, 1.6, 2.2 & 1.2, 1.6, 2.3 & 1.1, 1.7, 2.4 \\
10. NGC 6951 &  72 & 0.34957 & [$-$0.56, 0.13] & 0.95 & [0.83, 1.06] & 0.06 & 2.0, 2.7, 3.8 & 2.1, 2.9, 3.9 & 2.1, 2.9, 4.1 & 2.1, 3.0, 4.1 \\
11. NGC 1637 &  10 & 0.13771 & [$-$0.43, 0.63] & 0.94 & [0.63, 1.30] & 0.06 & 0.6, 0.9, 1.5 & 0.5, 1.0, 1.7 & 0.5, 1.0, 1.9 & 0.5, 1.0, 2.1 \\
12. NGC 3198 &   3 & 0.13795 & [$-$0.38, 0.66] & 0.86 & [0.50, 1.23] & 0.06 & 0.6, 1.2, 2.7 & 0.5, 1.4, 3.4 & 0.5, 1.5, 4.0 & 0.5, 1.5, 4.7 \\
13. NGC 4536 &   7 & 0.15383 & [$-$0.37, 0.68] & 0.88 & [0.57, 1.20] & 0.06 & 0.7, 1.1, 1.9 & 0.7, 1.2, 2.2 & 0.6, 1.4, 2.5 & 0.6, 1.3, 2.7 \\
\hline
{\bf Group Parameters}  & 957 & {\bf 0.04} & {\bf [$-$0.20, 0.27]} & {\bf 0.76} & {\bf [0.58, 0.94]} & 0.06 & 1.1, 2.4, 4.9 & 1.3, 2.8, 6.5 & 1.3, 3.1, 7.6 & 1.3, 3.3, 8.4 \\
\hline
\footnotetext[0]{$^1$ Entries indicate the 2.5\%, 50\%, and 97.5\%
  quantiles of \tdep\ (Gyr) at given values of \sigmol\ (\msunpc).}
\end{tabular}
\label{bayesgals}
\end{sidewaystable*}

\section*{Acknowledgements}

We thank Julia Roman Duval, Karl Gordon, Chris Hayward, J\'er\^ome
Pety, Greg Stinson, and Amelia Stutz for stimulating discussions on
star formation in the ISM, and the referee, Adam Leroy for providing a
thorough report that improved the paper.  The MCMC simulations were
run on the bwGRiD (http://www.bw-grid.de), member of the German D-Grid
initiative, funded by the Ministry for Education and Research
(Bundesministerium f\"ur Bildung und Forschung) and the Ministry for
Science, Research and Arts Baden-W\"urttemberg (Ministerium f\"ur
Wissenschaft, Forschung und Kunst Baden-W\"urttemberg).  RS, PCC, RSK,
and LKK acknowledge support from the Deutsche Forschungsgemeinschaft
(DFG) via the SFB 881 (B1 and B2) ``The Milky Way System,''
(subproject B1, B2, and B5) and the SPP (priority program) 1573
``Physics of the ISM''.  BK is supported from the Southern California
Center for Galaxy Evolution, a multi-campus research program funded by
the University of California Office of Research. NR acknowledges
support from South Africa Square Kilometer Array (SKA) Postdoctoral
Fellowship program.

\bibliography{citations}
\bibliographystyle{mn2e}

\label{lastpage}

\end{document}